# ROBUST ESTIMATES IN GENERALIZED PARTIALLY LINEAR MODELS[1]

BY GRACIELA BOENTE, XUMING HE AND JIANHUI ZHOU

*Universidad de Buenos Aires and CONICET, University of Illinois at Urbana-Champaign and University of Virginia*

In this paper, we introduce a family of robust estimates for the parametric and nonparametric components under a generalized partially linear model, where the data are modeled by $y_i|(\mathbf{x}_i, t_i) \sim F(\cdot, \mu_i)$ with $\mu_i = H(\eta(t_i) + \mathbf{x}_i^T \beta)$, for some known distribution function $F$ and link function $H$. It is shown that the estimates of $\beta$ are root-$n$ consistent and asymptotically normal. Through a Monte Carlo study, the performance of these estimators is compared with that of the classical ones.

**1. Introduction.** Semiparametric models contain both a parametric and a nonparametric component. Sometimes, the nonparametric component plays the role of a nuisance parameter. Much research has been done on estimators of the parametric component in a general framework, aiming to obtain asymptotically efficient estimators. The aim of this paper is to consider semiparametric versions of the generalized linear models where the response $y$ is to be predicted by covariates $(\mathbf{x}, t)$, where $\mathbf{x} \in \mathbb{R}^p$ and $t \in \mathcal{T} \subset \mathbb{R}$. It will be assumed that the conditional distribution of $y|(\mathbf{x}, t)$ belongs to the canonical exponential family $\exp[y\theta(\mathbf{x}, t) - B(\theta(\mathbf{x}, t)) + C(y)]$ for known functions $B$ and $C$. Then $\mu(\mathbf{x}, t) = \mathrm{E}(y|(\mathbf{x}, t)) = B'(\theta(\mathbf{x}, t))$, with $B'$ denoting the derivative of $B$. In generalized linear models [19], which constitute a popular approach for modeling a wide variety of data, it is often assumed that the mean is modeled linearly through a known inverse link function, $g$, that is,

$$g(\mu(\mathbf{x}, t)) = \beta_0 + \mathbf{x}^T \boldsymbol{\beta} + \alpha t.$$

Received January 2004; revised November 2006.
[1]Supported in part by Grants PICT 03-00000-006277 from ANPCYT, PID 5505 from CONICET, X-094 from the Universidad de Buenos Aires, 13900-6 from the Fundación Antorchas, Argentina and U.S. NSF Grant DMS-01-02411. The research began while the first author was granted with a Guggenheim fellowship.

For instance, an ordinary logistic regression model assumes that the observations $(y_i, \mathbf{x}_i, t_i)$ are such that the response variables are independent binomial variables $y_i|(\mathbf{x}_i, t_i) \sim Bi(1, p_i)$, whose success probabilities depend on the explanatory variables through the relation $g(p_i) = \beta_0 + \mathbf{x}_i^{\mathrm{T}} \boldsymbol{\beta} + \alpha t_i$, with $g(u) = \ln(u/(1-u))$.

The influence function of the classical estimates based on the quasi-likelihood is unbounded. Large deviations of the response from its mean, as measured by the Pearson residuals, or outlying points in the covariate space, can have a large influence on the estimators. Those outliers or potential outliers for the generalized linear regression model are to be detected and controlled by robust procedures such as those considered by Stefanski, Carroll and Ruppert [23], Künsch, Stefanski and Carroll [17], Bianco and Yohai [5] and Cantoni and Ronchetti [9].

In some applications, the linear model is insufficient to explain the relationship between the response variable and its associated covariates. To avoid the *curse of dimensionality*, we allow most predictors to be modeled linearly while a small number of predictors (possibly just one) enter the model nonparametrically. The relationship will be given by the semiparametric generalized partially linear model

$$(1) \qquad \mu(\mathbf{x}, t) = H(\eta(t) + \mathbf{x}^{\mathrm{T}} \boldsymbol{\beta}),$$

where $H = g^{-1}$ is a known link function, $\boldsymbol{\beta} \in \mathbb{R}^p$ is an unknown parameter and $\eta$ is an unknown continuous function.

Severini and Wong [22] introduced the concept of *generalized profile likelihood*, which was later applied to this model by Severini and Staniswalis [21]. In this method, the nonparametric component is viewed as a function of the parametric component and $\sqrt{n}$-consistent estimates for the parametric component can be obtained when the usual optimal rate for the smoothing parameter is used. Such estimates do not deal with outlying observations. In a semiparametric setting, outliers can have a devastating effect, since the extreme points can easily affect the scale and shape of the function estimate of $\eta$, leading to possible wrong conclusions concerning $\beta$. The basic ideas from robust smoothing and robust regression estimation have been adapted to partially linear regression models where $H(t) = t$; we refer to [3, 13, 15]. A robust generalized estimating equations approach for generalized partially linear models with clustered data, using regression splines and Pearson residuals, is given in [14].

In Section 2 of the present paper, we introduce a two-step robust procedure for estimating the parameter $\boldsymbol{\beta}$ and the function $\eta$ under the generalized partially linear model (1). In Section 3, we give conditions under which the proposed method will lead to strongly consistent estimators and in Section 4, we derive the asymptotic distribution of those estimators. In Section 5, simulation studies are carried out to assess the robustness and efficiency of the proposals. The proofs are deferred to the Appendix.

**2. The proposal.**

2.1. *The estimators.* Let $(y_i, \mathbf{x}_i, t_i)$ be independent observations such that $y_i|(\mathbf{x}_i, t_i) \sim F(\cdot, \mu_i)$, with $\mu_i = H(\eta(t_i) + \mathbf{x}_i^{\mathrm{T}}\boldsymbol{\beta})$ and $\mathrm{VAR}(y_i|(\mathbf{x}_i, t_i)) = V(\mu_i)$. Let $\eta_0(t)$ and $\boldsymbol{\beta}_0$ denote the true parameter values and $\mathrm{E}_0$ the expected value under the true model, so that $\mathrm{E}_0(y|(\mathbf{x},t)) = H(\eta_0(t) + \mathbf{x}^{\mathrm{T}}\boldsymbol{\beta}_0)$. Letting $\rho(y, u)$ be a loss function to be specified in the next subsection, we define

$$(2) \qquad S_n(a, \boldsymbol{\beta}, t) = \sum_{i=1}^{n} W_i(t) \rho(y_i, \mathbf{x}_i^{\mathrm{T}} \boldsymbol{\beta} + a) w_1(\mathbf{x}_i),$$

$$(3) \qquad S(a, \boldsymbol{\beta}, \tau) = \mathrm{E}_0[\rho(y, \mathbf{x}^{\mathrm{T}} \boldsymbol{\beta} + a) w_1(\mathbf{x}) | t = \tau],$$

where $W_i(t)$ are the kernel (or nearest-neighbor with kernel) weights on $t_i$ and $w_1(\cdot)$ is a function that downweights high leverage points in the $\mathbf{x}$ space. Note that $S_n(a, \boldsymbol{\beta}, \tau)$ is an estimate of $S(a, \boldsymbol{\beta}, \tau)$, which is a continuous function of $(a, \boldsymbol{\beta}, \tau)$ if $(y, \mathbf{x})|t = \tau$ has a distribution function that is continuous with respect to $\tau$.

Fisher consistency states that $\eta_0(t) = \mathrm{argmin}_a S(a, \boldsymbol{\beta}_0, t)$. This is a key point in order to get asymptotically unbiased estimates for the nonparametric component. In many situations, a stronger condition holds, that is, under general conditions, it can be verified that

$$(4) \qquad S(\eta_0(t), \boldsymbol{\beta}_0, t) < S(a, \boldsymbol{\beta}, t) \qquad \forall \boldsymbol{\beta} \neq \boldsymbol{\beta}_0, \ a \neq \eta_0(t),$$

which entails Fisher consistency.

Following the ideas of Severini and Staniswalis [21], we define the function $\eta_{\boldsymbol{\beta}}(t)$ as the minimizer of $S(a, \boldsymbol{\beta}, t)$ that will be estimated by the minimizer $\hat{\eta}_{\boldsymbol{\beta}}(t)$ of $S_n(a, \boldsymbol{\beta}, t)$.

To provide an estimate of $\boldsymbol{\beta}$ with root-$n$ convergence rate, we denote

$$(5) \qquad F_n(\boldsymbol{\beta}) = n^{-1} \sum_{i=1}^{n} \rho(y_i, \mathbf{x}_i^{\mathrm{T}} \boldsymbol{\beta} + \hat{\eta}_{\boldsymbol{\beta}}(t_i)) w_2(\mathbf{x}_i),$$

$$(6) \qquad F(\boldsymbol{\beta}) = \mathrm{E}_0[\rho(y, \mathbf{x}^{\mathrm{T}} \boldsymbol{\beta} + \eta_{\boldsymbol{\beta}}(t)) w_2(\mathbf{x})],$$

where $w_2(\cdot)$ plays the same role (and can be taken to be the same) as $w_1(\cdot)$. We will assume that $\boldsymbol{\beta}_0$ is the unique minimizer of $F(\boldsymbol{\beta})$. This assumption is a standard condition in M-estimation in order to get consistent estimators of the parametric component and is analogous to condition (A-4) of [16], page 129.

A two-step robust proposal is now given as follows:

- **Step 1**: For each value of $t$ and $\boldsymbol{\beta}$, let

$$(7) \qquad \hat{\eta}_{\boldsymbol{\beta}}(t) = \underset{a \in \mathbb{R}}{\mathrm{argmin}}\, S_n(a, \boldsymbol{\beta}, t).$$

- **Step 2**: Define the estimate of $\boldsymbol{\beta}_0$ as

(8) $$\hat{\boldsymbol{\beta}} = \operatorname*{argmin}_{\boldsymbol{\beta} \in \mathbb{R}^p} F_n(\boldsymbol{\beta})$$

and the estimate of $\eta_0(t)$ as $\hat{\eta}_{\hat{\boldsymbol{\beta}}}(t)$.

2.2. *Loss function $\rho$.* We propose two classes of loss functions. The first aims to bound the deviances, while the second, introduced by Cantoni and Ronchetti [9], bounds the Pearson residuals.

The first class of loss functions takes the form

(9) $$\rho(y, u) = \phi[-\ln f(y, H(u)) + A(y)] + G(H(u)),$$

where $\phi$ is a bounded nondecreasing function with continuous derivative $\varphi$ and $f(\cdot, s)$ is the density of the distribution function $F(\cdot, s)$ with $y|(\mathbf{x}, t) \sim F(\cdot, H(\eta_0(t) + \mathbf{x}^{\mathrm{T}} \boldsymbol{\beta}_0))$. To avoid triviality, we also assume that $\phi$ is nonconstant in a positive probability set. Typically, $\phi$ is a function which behaves like the identity function in a neighborhood of 0. The function $A(y)$ is typically used to remove a term from the log-likelihood that is independent of the parameter and can be defined as $A(y) = \ln(f(y, y))$ in order to obtain the deviance. The correction term $G$ is used to guarantee Fisher consistency and satisfies

$$G'(s) = \int \varphi[-\ln f(y, s) + A(y)] f'(y, s) \, d\mu(y)$$
$$= \mathrm{E}_s(\varphi[-\ln f(y, s) + A(y)] f'(y, s)/f(y, s)),$$

where $\mathrm{E}_s$ indicates expectation taken under $y \sim F(\cdot, s)$ and $f'(y, s)$ is shorthand for $\partial f(y, s)/\partial s$. With this class of $\rho$ functions, we call the resulting estimator a *modified likelihood estimator*.

In a logistic regression setting, Bianco and Yohai [5] considered the score function

$$\phi(t) = \begin{cases} t - t^2/2c & \text{if } t \leq c, \\ c/2 & \text{otherwise}, \end{cases}$$

while Croux and Haesbroeck [12] proposed using the score function

$$\phi(t) = \begin{cases} t \exp(-\sqrt{c}) & \text{if } t \leq c, \\ -2(1 + \sqrt{t}) \exp(-\sqrt{t}) + (2(1 + \sqrt{c}) + c) \exp(-\sqrt{c}) & \text{otherwise}. \end{cases}$$

Both score functions can be used in the general setting. Explicit forms of the correction term $G(s)$ for the binomial and Poisson families are given in [1]. It is worth noting that when considering the deviance and a continuous family of distributions with strongly unimodal density function, the correction term $G$ can be avoided, as discussed in [4].

The second class of loss functions is based on [9], wherein the authors consider a general class of M-estimators of Mallows type by separately bounding the influence of deviations on $y$ and $(\mathbf{x}, t)$. Their approach is based on robustifying the quasi-likelihood, which is an alternative to the generalizations given for generalized linear regression models by Stefanski, Carroll and Ruppert [23] and Künsch, Stefanski and Carroll [17]. Let $r(y, \mu) = (y - \mu) V^{-1/2}(\mu)$ be the Pearson residuals with $\text{VAR}(y_i | (\mathbf{x}_i, t_i)) = V(\mu_i)$. Denote $\nu(y, \mu) = V^{-1/2}(\mu) \psi_c(r(y, \mu))$, where $\psi_c$ is an odd nondecreasing score function with tuning constant $c$, such as the Huber function and

$$(10) \qquad \rho(y, u) = -\bigg[ \int_{s_0}^{H(u)} \nu(y, s)\, ds + G(H(u)) \bigg],$$

where $s_0$ is such that $\nu(y, s_0) = 0$ and the correction term (included to ensure Fisher consistency), also denoted $G(s)$, satisfies $G'(s) = -\text{E}_s(\nu(y, s))$. With such a $\rho$ function, we call the resulting estimator a *robust quasi-likelihood estimator*. For the binomial and Poisson families, explicit forms of the correction term $G(s)$ are given in [9].

### 2.3. General comments.

(a) *Fisher consistency and uniqueness.* Under a logistic partially linear regression model, if

$$(11) \qquad P(\mathbf{x}^{\mathrm{T}} \boldsymbol{\beta} = \alpha | t = \tau) < 1, \qquad \forall (\boldsymbol{\beta}, \alpha) \neq 0 \text{ and } \tau \in \mathcal{T},$$

and if we consider the loss function given by (9) with $\phi$ satisfying the regularity conditions given in [5], it is easy to see that (4) holds and that Fisher consistency for the nonparametric component is attained under this model. Moreover, it is easy to verify that $\boldsymbol{\beta}_0$ is the unique minimizer of $F(\boldsymbol{\beta})$ in this case. The same assertion can be verified for the robust quasi-likelihood proposal if $\psi_c$ is bounded and increasing.

Under a generalized partially linear model with the response having a gamma distribution with a fixed shape parameter, Theorem 1 of Bianco, García Ben and Yohai [4] allows us to verify (4) and Fisher consistency for the nonparametric and parametric components if the score function $\phi$ is bounded and strictly increasing on the set where it is not constant and if (11) holds.

For any generalized partially linear model, conditions similar to those considered in [9] will lead to the desired uniqueness implied by (4). Note that this condition is quite similar to Condition (E) of [21], page 511. When considering the classical quasi-likelihood, the assumption $\boldsymbol{\beta}_0 = \operatorname{argmin}_{\boldsymbol{\beta}} F(\boldsymbol{\beta})$ is related to Condition (7.e) of [21], page 511, but for the robust quasi-likelihood, this assumption is fulfilled, for instance, for a gamma family with a fixed shape parameter such that (11) holds and $\psi_c$ is bounded and increasing.

(b) *Differentiated equations.* If the function $\rho(y, u)$ is continuously differentiable and we denote $\Psi(y, u) = (\partial \rho(y, u))/\partial u$, the estimates will be solutions to the differentiated equations. More precisely, $\eta_{\boldsymbol{\beta}}(t)$ and $\hat{\eta}_{\boldsymbol{\beta}}(t)$ will be solutions to $S^1(a, \boldsymbol{\beta}, t) = 0$ and $S_n^1(a, \boldsymbol{\beta}, t) = 0$, respectively, with

$$S^1(a, \boldsymbol{\beta}, \tau) = E(\Psi(y, \mathbf{x}^{\mathrm{T}} \boldsymbol{\beta} + a) w_1(\mathbf{x}) | t = \tau), \tag{12}$$

$$S_n^1(a, \boldsymbol{\beta}, t) = \sum_{i=1}^{n} W_i(t) \Psi(y_i, \mathbf{x}_i^{\mathrm{T}} \boldsymbol{\beta} + a) w_1(\mathbf{x}_i). \tag{13}$$

Furthermore, $\hat{\boldsymbol{\beta}}$ is a solution of $F_n^1(\boldsymbol{\beta}) = 0$ and Fisher consistency implies that $F^1(\boldsymbol{\beta}_0) = 0$ and $S^1(\eta_0(t), \boldsymbol{\beta}_0, t) = 0$, where

$$F^1(\boldsymbol{\beta}) = E\left(\Psi(y, \mathbf{x}^{\mathrm{T}} \boldsymbol{\beta} + \eta_{\boldsymbol{\beta}}(t)) w_2(\mathbf{x}) \left[\mathbf{x} + \frac{\partial}{\partial \boldsymbol{\beta}} \eta_{\boldsymbol{\beta}}(t)\right]\right), \tag{14}$$

$$F_n^1(\boldsymbol{\beta}) = n^{-1} \sum_{i=1}^{n} \Psi(y_i, \mathbf{x}_i^{\mathrm{T}} \boldsymbol{\beta} + \hat{\eta}_{\boldsymbol{\beta}}(t_i)) w_2(\mathbf{x}_i) \left[\mathbf{x}_i + \frac{\partial}{\partial \boldsymbol{\beta}} \hat{\eta}_{\boldsymbol{\beta}}(t_i)\right]. \tag{15}$$

Note that these first order equations may have multiple solutions and, therefore, we may need the values of the objective functions (2) and (5) to select the final estimator. For a family of distributions with positive and finite information number, Bianco and Boente [1] give conditions that entail the following: for each $t$, there exists a neighborhood of $\eta_0(t)$ where $S^1(\eta_0(t), \boldsymbol{\beta}_0, t) = 0$ and $S^1(a, \boldsymbol{\beta}_0, t) \neq 0$ for $a \neq \eta_0(t)$. Moreover, $\eta_0(t)$ corresponds to a local minimum of $S(a, \boldsymbol{\beta}_0, t)$. The asymptotic results in this paper are derived by assuming the existence of a unique minimum; otherwise, one can only ensure that there exists a solution to the estimating equations that is consistent.

In the modified likelihood approach, the derivative of (9) is given by $\Psi(y, u) = H'(u)[\Psi_1(y, H(u)) + G'(H(u))]$, where

$$\Psi_1(y, u) = \varphi[-\ln f(y, H(u)) + A(y)][-f'(y, H(u))/f(y, H(u))].$$

On the other hand, for the proposal based on the robust quasi-likelihood, we have the following expression for the derivative of (10):

$$\begin{aligned} \Psi(y, u) &= -[\nu(y, H(u)) + G'(H(u))] H'(u) \\ &= -[\psi_c(r(y, H(u))) V^{-1/2}(H(u)) + G'(H(u))] H'(u) \\ &= -[\psi_c(r(y, H(u))) - E_{H(u)}\{\psi_c(r(y, H(u)))\}] H'(u) V^{-1/2}(H(u)). \end{aligned}$$

One advantage of solving $S_n^1(a, \boldsymbol{\beta}, t) = 0$ and $F_n^1(\boldsymbol{\beta}) = 0$ is to avoid the numerical integration involved in the loss function (10), but the uniqueness of the solutions might be difficult to guarantee in general, except for those cases discussed in part (a) of this section. Also, note that when using the score function of Croux and Haesbroeck [12], the function $G(s)$ in (9) has an explicit expression which does not require any numerical integration.

(c) *Some robustness issues.* It is clear that for unbounded response variables $y$, a bounded score function allows us to deal with large residuals. For models with a bounded response, for example, under a logistic model, the advantage of a bounded score function is mainly to guard against outliers with large Pearson residuals. If a binary response $y$ is contaminated, the Pearson residuals are large only when the variances at the contaminated points are close to 0. These points are made more specific in the simulation study in Section 5.

It is also worth noting that our robust procedures are effective only if at least one nonconstant covariate $\mathbf{x}$ is present. To consider a case without any covariate, we may take $y_i \sim Bi(1,p)$ as a random sample. Then easy calculations show that the minimizer $\hat{a}$ of $S_n(a) = n^{-1} \sum_{i=1}^n \rho(y_i, a)$ equals the classical estimator, that is, $\hat{a} = H^{-1}(\sum_{i=1}^n y_i/n)$ with $H(u) = 1/(1+\exp(-u))$, when using either the score function proposed in [5] or that given by Cantoni and Ronchetti [9]. The same situation obtains if $y_i|t_i \sim Bi(1, p(t_i))$, where the resulting estimate of $p(t)$ will be the local mean. In the present paper, with a semiparametric model where the covariate $\mathbf{x}$ plays a role, both downweighting the leverage points and controlling outlying responses work toward robustness.

**3. Consistency.** We will assume that $t \in \mathcal{T}$ and let $\mathcal{T}_0 \subset \mathcal{T}$ be a compact set. For any continuous function $v : \mathcal{T} \to \mathbb{R}$, we will denote $\|v\|_\infty = \sup_{t \in \mathcal{T}} |v(t)|$ and $\|v\|_{0,\infty} = \sup_{t \in \mathcal{T}_0} |v(t)|$.

In this section, we will show that the estimates defined by means of (7) and (8) are consistent under mild conditions, when the smoother weights are the kernel weights $W_i(t) = (\sum_{j=1}^n K((t-t_j)/h_n))^{-1} K((t-t_i)/h_n)$. Analogous results can be obtained for the weights based on nearest neighbors using arguments similar to those considered in [6]. In this paper, we will use the following set of assumptions:

C1. The function $\rho(y,a)$ is continuous and bounded and the functions $\Psi(y,a) = \partial \rho(y,a)/\partial a$, $w_1(.)$ and $w_2(.)$ are bounded.
C2. The kernel $K : \mathbb{R} \to \mathbb{R}$ is an even, nonnegative, continuous and bounded function, satisfying $\int K(u)\,du = 1$, $\int u^2 K(u)\,du < \infty$ and $|u|K(u) \to 0$ as $|u| \to \infty$.
C3. The bandwidth sequence $h_n$ is such that $h_n \to 0$ and $nh_n/\log(n) \to \infty$.
C4. The marginal density $f_T$ of $t$ is a bounded function and given any compact set $\mathcal{T}_0 \subset \mathcal{T}$, there exists a positive constant $A_1(\mathcal{T}_0)$ such that $A_1(\mathcal{T}_0) < f_T(t)$ for all $t \in \mathcal{T}_0$.
C5. The function $S(a, \boldsymbol{\beta}, t)$ satisfies the following equicontinuity condition: for any $\varepsilon > 0$, there exists some $\delta > 0$ such that for any $t_1, t_2 \in \mathcal{T}_0$ and $\boldsymbol{\beta}_1, \boldsymbol{\beta}_2 \in \mathcal{K}$, a compact set in $R^p$,

$$|t_1 - t_2| < \delta \text{ and } \|\boldsymbol{\beta}_1 - \boldsymbol{\beta}_2\| < \delta \Rightarrow \sup_{a \in \mathbb{R}} |S(a, \boldsymbol{\beta}_1, t_1) - S(a, \boldsymbol{\beta}_2, t_2)| < \varepsilon.$$

C6. The function $S(a,\boldsymbol{\beta},t)$ is continuous and $\eta_{\boldsymbol{\beta}}(t)$ is a continuous function of $(\boldsymbol{\beta},t)$.

REMARK 3.1. If the conditional distribution of $\mathbf{x}|t=\tau$ is continuous with respect to $\tau$, the continuity and boundness of $\rho$ stated in C1 entail that $S(a,\boldsymbol{\beta},\tau)$ is continuous.

Assumption C3 ensures that for each fixed $a$ and $\boldsymbol{\beta}$, we have convergence of the kernel estimates to their mean, while C5 guarantees that the bias term converges to 0.

Assumption C4 is a standard condition in semiparametric models. In the classical case, it corresponds to condition (D) of [21], page 511. It is also considered in nonparametric regression when the uniform consistency results on the $t$-space are needed; it allows us to deal with the denominator in the definition of the kernel weights, which is, in fact, an estimate of the marginal density $f_T$.

Assumption C5 is fulfilled under C1 if the following equicontinuity condition holds: for any $\varepsilon > 0$, there exist compact sets $\mathcal{K}_1 \subset \mathbb{R}$ and $\mathcal{K}_p \subset \mathbb{R}^p$ such that for any $\tau \in \mathcal{T}_0$, $P((y,\mathbf{x}) \in \mathcal{K}_1 \times \mathcal{K}_p|t=\tau) > 1 - \varepsilon$, which holds, for instance, if, for $1 \leq i \leq n$ and $1 \leq j \leq p$, $x_{ij} = \phi_j(t_i) + u_{ij}$, where $\phi_j$ are continuous functions and $u_{ij}$ are i.i.d and independent of $t_i$.

THEOREM 3.1. *Let $\mathcal{K} \subset \mathbb{R}^p$ and $\mathcal{T}_0 \subset \mathcal{T}$ be compact sets such that $\mathcal{T}_\delta \subset \mathcal{T}$, where $\mathcal{T}_\delta$ is the closure of a $\delta$-neighborhood of $\mathcal{T}_0$. Assume that C1–C6 and the following conditions hold:*

(i) *$K$ is of bounded variation;*
(ii) *the family of functions $\mathcal{F} = \{f(y,\mathbf{x}) = \rho(y,\mathbf{x}^{\mathrm{T}}\boldsymbol{\beta}+a)w_1(\mathbf{x}),\ \boldsymbol{\beta} \in \mathcal{K},\ a \in \mathbb{R}\}$ has covering number $N(\varepsilon,\mathcal{F},L^1(\mathbb{Q})) \leq A\varepsilon^{-W}$, for any probability $\mathbb{Q}$ and $0 < \varepsilon < 1$.*

*Then we have*

(a) $\sup_{\substack{\boldsymbol{\beta} \in \mathcal{K} \\ a \in \mathbb{R}}} \|S_n(a,\boldsymbol{\beta},\cdot) - S(a,\boldsymbol{\beta},\cdot)\|_{0,\infty} \xrightarrow{a.s.} 0$;

(b) *if $\inf_{\substack{\boldsymbol{\beta} \in \mathcal{K} \\ t \in \mathcal{T}_0}}[\lim_{|a| \to \infty} S(a,\boldsymbol{\beta},t) - S(\eta_{\boldsymbol{\beta}}(t),\boldsymbol{\beta},t)] > 0$, then*

$$\sup_{\boldsymbol{\beta} \in \mathcal{K}} \|\hat{\eta}_{\boldsymbol{\beta}} - \eta_{\boldsymbol{\beta}}\|_{0,\infty} \xrightarrow{a.s.} 0.$$

THEOREM 3.2. *Let $\hat{\boldsymbol{\beta}}$ be the minimizer of $F_n(\boldsymbol{\beta})$, where $F_n(\boldsymbol{\beta})$ is defined as in* (5), *with $\hat{\eta}_{\boldsymbol{\beta}}$ satisfying*

(16) $$\sup_{\boldsymbol{\beta} \in \mathcal{K}} \|\hat{\eta}_{\boldsymbol{\beta}} - \eta_{\boldsymbol{\beta}}\|_{0,\infty} \xrightarrow{a.s.} 0$$

*for any compact set $\mathcal{K}$ in $R^p$. If C1 holds, then*

(a) $\sup_{\boldsymbol{\beta}\in\mathcal{K}}|F_n(\boldsymbol{\beta}) - F(\boldsymbol{\beta})| \xrightarrow{a.s.} 0$;

(b) *if, in addition, there exists a compact set $\mathcal{K}_1$ such that* $\lim_{m\to\infty} P(\bigcap_{n\geq m} \hat{\boldsymbol{\beta}} \in \mathcal{K}_1) = 1$ *and $F(\boldsymbol{\beta})$ has a unique minimum at $\boldsymbol{\beta}_0$, then* $\hat{\boldsymbol{\beta}} \xrightarrow{a.s.} \boldsymbol{\beta}_0$.

REMARK 3.2. Theorems 3.1 and 3.2 entail that $\|\hat{\eta}_{\hat{\boldsymbol{\beta}}} - \eta_0\|_{0,\infty} \xrightarrow{a.s.} 0$, since $\eta_{\boldsymbol{\beta}}(t)$ is continuous. For the covering number used in Condition (ii) of Theorem 3.1, see [20].

**4. Asymptotic normality.** From now on, $\mathcal{T}$ is assumed to be a compact set. A set of assumptions denoted N1–N6, under which the resulting estimates are asymptotically normally distributed, are detailed in the Appendix.

THEOREM 4.1. *Assume that the $t_i$'s are random variables with distribution on a compact set $\mathcal{T}$ and that N1–N6 hold. Then for any consistent solution $\hat{\boldsymbol{\beta}}$ of (15), we have*

$$\sqrt{n}(\hat{\boldsymbol{\beta}} - \boldsymbol{\beta}_0) \xrightarrow{D} N(\mathbf{0}, \mathbf{A}^{-1}\boldsymbol{\Sigma}(\mathbf{A}^{-1})^{\mathrm{T}}),$$

*where $\mathbf{A}$ is defined in N3 and $\boldsymbol{\Sigma}$ is defined in N4.*

REMARK 4.1. Theorem 4.1 can be used to construct a Wald-type statistic to make inferences involving only a subset of the regression parameter, that is, when we want to test $H_0 : \boldsymbol{\beta}_{(2)} = \mathbf{0}$, with $\boldsymbol{\beta}^{\mathrm{T}} = (\boldsymbol{\beta}_{(1)}^{\mathrm{T}}, \boldsymbol{\beta}_{(2)}^{\mathrm{T}})$.

Likelihood ratio-type tests can also be used based on the robust quasi-likelihood introduced in Section 2, as was done for generalized linear models by Cantoni and Ronchetti [9], or on the robustified deviance. A robust measure of discrepancy between the two models is defined as

$$\Lambda = 2\left[\sum_{i=1}^{n} \rho(y_i, \mathbf{x}_i^{\mathrm{T}}\hat{\boldsymbol{\beta}} + \hat{\eta}_{\hat{\boldsymbol{\beta}}}(t_i))w_2(\mathbf{x}_i) - \sum_{i=1}^{n} \rho(y_i, \mathbf{x}_i^{\mathrm{T}}\hat{\boldsymbol{\beta}}_0 + \hat{\eta}_{\hat{\boldsymbol{\beta}}_0}(t_i))w_2(\mathbf{x}_i)\right],$$

where $\hat{\boldsymbol{\beta}}_0^{\mathrm{T}} = (\hat{\boldsymbol{\beta}}_{(1)}^{\mathrm{T}}, \mathbf{0}^{\mathrm{T}})$ is the estimate of $\boldsymbol{\beta}$ under the null hypothesis. Both estimates $\hat{\boldsymbol{\beta}}_0$ and $\hat{\boldsymbol{\beta}}$ need to be computed using the same loss function $\rho$ considered in $\Lambda$, in order to ensure that $\Lambda$ will behave asymptotically as a linear combination of independent chi-square random variables with one degree of freedom. As in [9], it can be seen that $\Lambda = n\mathbf{U}_{n,(2)}^{\mathrm{T}}\mathbf{A}_{22.1}\mathbf{U}_{n,(2)} + o_p(1)$, with $\mathbf{A}_{22.1} = \mathbf{A}_{22} - \mathbf{A}_{21}\mathbf{A}_{11}^{-1}\mathbf{A}_{12}$ and $\sqrt{n}\mathbf{U}_n \xrightarrow{D} N(\mathbf{0}, \mathbf{A}^{-1}\boldsymbol{\Sigma}(\mathbf{A}^{-1})^{\mathrm{T}})$.

**5. Monte Carlo study.** A small-scale simulation study was carried out to assess the performance of the robust estimators considered in this paper. A one-dimensional covariate $x$ and a nonparametric function $\eta(t)$ were considered. The modified likelihood estimator (MOD) used the score function of Croux and Haesbroeck [12] with $c = 0.5$. With this choice, the function $G(s)$ has an explicit expression, so no numerical integration is necessary. The weight functions take the form

$$w_1^2(x_i) = w_2^2(x_i) = \{1 + (x_i - M_n)^2\}^{-1},$$

where $M_n = Median\{x_j : j = 1, \ldots, n\}$ is the sample median.

The two competitors considered in the study were the quasi-likelihood estimator (QAL) of Severeni and Staniswalis [21] and the robust quasi-likelihood estimator (RQL) of Cantoni and Ronchetti [9]. For the latter, the Huber function $\psi_c(x) = \max\{-1.2, \min(1.2, x)\}$ was used with the same weight functions as above. The QAL estimator corresponds to $\psi_c(x) = x$ and $w_1(x) = w_2(x) = 1$. In all cases, the kernel $K(t) = \max\{0, 1 - |t|\}$ was used. In Studies 1 and 3 below, the search for $\boldsymbol{\beta}$ uses a grid of size 0.05, while in Study 2 the grid size is 0.01.

An important issue in any smoothing procedure is the choice of the smoothing parameter. Under a nonparametric regression model with $\boldsymbol{\beta} = 0$ and $H(t) = t$, two commonly used approaches are cross-validation and plug-in. However, these procedures may not be robust; their sensitivity to anomalous data was discussed by several authors, including [7, 10, 18, 24]. Wang and Scott [24] note that in the presence of outliers, the least squares cross-validation function is nearly constant on its whole domain and, thus, essentially worthless for the purpose of choosing a bandwidth. The robustness issue remains for the estimators considered in this paper. With a small bandwidth, a small number of outliers with similar values of $t_i$ could easily drive the estimate of $\eta$ to dangerous levels. Therefore, we may consider a robust cross-validation approach as follows:

- Select at random a subset of size $100(1-\alpha)\%$. Let $\mathcal{I}_{1-\alpha}$ denote the indexes of these observations and $\mathcal{J}_{1-\alpha}$ the indexes of the remaining ones.
- For each given $h$, compute

$$\hat{\eta}_{\boldsymbol{\beta}}^{(-\alpha)}(t,h) = \operatorname*{argmin}_{a \in \mathbb{R}} \sum_{i \in \mathcal{I}_{1-\alpha}} W_i(t,h) \rho(y_i, \mathbf{x}_i^{\mathrm{T}} \boldsymbol{\beta} + a) w_1(\mathbf{x}_i),$$

$$\hat{\boldsymbol{\beta}}^{(-\alpha)}(h) = \operatorname*{argmin}_{\boldsymbol{\beta} \in \mathbb{R}^p} \sum_{i \in \mathcal{I}_{1-\alpha}} \rho(y_i, \mathbf{x}_i^{\mathrm{T}} \boldsymbol{\beta} + \hat{\eta}_{\boldsymbol{\beta}}^{(-\alpha)}(t_i, h)) w_2(\mathbf{x}_i),$$

where $W_i(t,h) = \{\sum_{j=1}^n K((t-t_j)/h)\}^{-1} K((t-t_i)/h)$.
- Choose

$$\hat{h}_n = \operatorname*{argmin}_{h} \sum_{i \in \mathcal{J}_{1-\alpha}} \rho(y_i, \mathbf{x}_i^{\mathrm{T}} \hat{\boldsymbol{\beta}}^{(-\alpha)}(h) + \hat{\eta}_{\hat{\boldsymbol{\beta}}^{(-\alpha)}}^{(-\alpha)}(t_i, h)) w_2(\mathbf{x}_i).$$

When the sample size $n$ is small, the leave-one-out cross-validation, which is similar to the approach considered here, is usually preferred. When $n$ is modestly large, the $v$-fold cross-validation is often used. However, both of them are computationally expensive. Based on our experience with a number of data sets, including some from Study 1 below, we found that the approach considered here is helpful. A full evaluation of this approach has not yet been completed.

To measure performance through simulation, we use the bias and standard deviation for the $\boldsymbol{\beta}$ estimate as well as the mean square error of the function estimate

$$\text{MSE}(\hat{\eta}) = n^{-1} \sum_{i=1}^{n} [\hat{\eta}(t_i) - \eta(t_i)]^2.$$

We report the comparisons in three scenarios as follows.

*Study* 1. Random samples of size $n = 100$ were generated from the model

$$x \sim \mathcal{U}(-1,1), \qquad t \sim \mathcal{U}(\{0.1, 0.2, \ldots, 1.0\}), \qquad y|(x,t) \sim Bi(10, p(x,t)),$$

where $\log(p(x,t)/(1 - p(x,t))) = 3x + e^{2t} - 4$. We summarized the results over 100 runs in Table 1, using three different bandwidths, $h_n = 0.1$, $h_n = 0.2$ and $h_n = 0.3$. The three estimates are labeled as $\text{QAL}(h_n)$, $\text{RQL}(h_n)$ and $\text{MOD}(h_n)$. Figure 1 gives the histograms of the estimates of $\beta$ for each method and bandwidth. It is clear that the robust estimators RQL and MOD have similar performance and that the relative efficiencies of the $\text{MOD}(h_n)$ are between 0.69 and 0.80, as compared to QAL $(h_n)$. The MOD method tends to have smaller bias than the RQL method and even than the QAL method. The normality of $\hat{\hat{\beta}}$ appeared to hold up quite well at this sample size.

TABLE 1
*Summary results for Study 1*

|  | Bias($\hat{\beta}$) | SD($\hat{\beta}$) | MSE($\hat{\beta}$) | MSE($\hat{\eta}$) |
|---|---|---|---|---|
| QAL(0.1) | 0.059 | 0.219 | 0.051 | 0.111 |
| QAL(0.2) | 0.033 | 0.214 | 0.047 | 0.073 |
| QAL(0.3) | 0.004 | 0.220 | 0.048 | 0.152 |
| RQL(0.1) | −0.051 | 0.242 | 0.061 | 0.114 |
| RQL(0.2) | −0.054 | 0.254 | 0.067 | 0.089 |
| RQL(0.3) | −0.105 | 0.262 | 0.080 | 0.154 |
| MOD(0.1) | 0.030 | 0.252 | 0.064 | 0.143 |
| MOD(0.2) | 0.018 | 0.251 | 0.063 | 0.088 |
| MOD(0.3) | −0.001 | 0.252 | 0.064 | 0.135 |

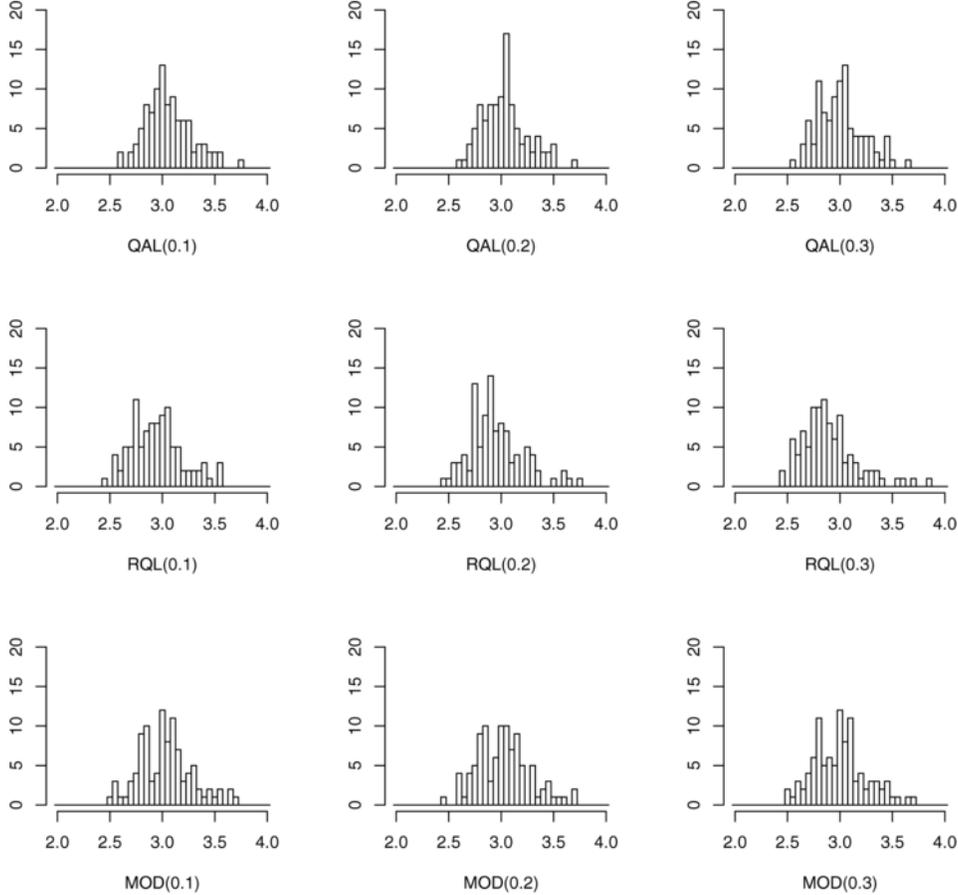

FIG. 1. *Histograms of $\hat{\boldsymbol{\beta}}$ for QAL, RQL and MOD using bandwidths $h_n = 0.1$, 0.2 and 0.3.*

We also applied the data-adaptive method described in this section for choosing $h_n$ based on a split of the sample into a training set (80% of the data) and a validation set (20%). On a total of ten random samples for Study 1, the resulting $h_n$ are mostly between 0.1 and 0.2. From Table 1, we may observe that $h_n = 0.2$ is indeed a good choice, but the performance of $\hat{\beta}$ is not very sensitive to the choice of $h_n$.

*Study* 2. To see how the robust estimators protect us from gross errors in the data, we generated a data set of size $n = 100$ from the model

$$x \sim N(0,1), \qquad t \sim N(1/2, 1/6), \qquad y|(x,t) \sim Bi(10, p(x,t)),$$

where $\log(p(x,t)/(1-p(x,t))) = 2x + 0.2$. We then replaced the first one, two and three observations by gross outliers. Table 2 gives the parameter estimates under the contaminated data, with $h_n = 0.1$, where $(x_i, y_i)$, $1 \leq i \leq 3$, denote the outliers. It is clear that the QAL estimate of $\beta$ was very sensitive to a single outlier, whereas the robust estimators remained stable.

TABLE 2
*Estimates of $\beta$ (true value of 2) in Study 2. $(x_i, y_i)$, $1 \leq i \leq 3$, denote the three contaminating points which replace the first three observations one by one*

|  | QAL | RQL | MOD |
|---|---|---|---|
| Original data | 2.02 | 2.08 | 1.99 |
| $x_1 = 10, y_1 = 0$ | 0.90 | 2.07 | 2.00 |
| $x_2 = -10, y_2 = 10$ | 0.31 | 2.06 | 1.97 |
| $x_3 = -10, y_3 = 10$ | 0.12 | 2.05 | 1.95 |

TABLE 3
*Summary results for Study 3*

| Data | Estimator | Bias($\hat{\beta}$) | SD($\hat{\beta}$) | MSE($\hat{\beta}$) | MSE($\hat{\eta}$) |
|---|---|---|---|---|---|
| Original | QAL | 0.126 | 0.357 | 0.143 | 0.297 |
| Original | RQL | 0.199 | 0.409 | 0.207 | 0.348 |
| Original | MOD | 0.158 | 0.386 | 0.174 | 0.317 |
| Contaminated $C_1$ | QAL | $-0.393$ | 0.366 | 0.288 | 0.378 |
| Contaminated $C_1$ | RQL | $-0.171$ | 0.440 | 0.223 | 0.378 |
| Contaminated $C_1$ | MOD | $-0.245$ | 0.414 | 0.231 | 0.365 |
| Contaminated $C_2$ | QAL | $-0.935$ | 0.287 | 0.957 | 0.446 |
| Contaminated $C_2$ | RQL | 0.018 | 0.545 | 0.297 | 0.399 |
| Contaminated $C_2$ | MOD | $-0.237$ | 0.436 | 0.246 | 0.350 |
| Contaminated $C_3$ | QAL | $-2.187$ | 0.071 | 4.788 | 0.402 |
| Contaminated $C_3$ | RQL | 0.177 | 0.430 | 0.216 | 0.400 |
| Contaminated $C_3$ | MOD | $-0.037$ | 0.475 | 0.227 | 0.369 |

*Study* 3. We considered data sets of size $n = 200$ which are generated from a bivariate normal distribution $(x_i, t_i) \sim N((0, 1/2), \Sigma)$, truncated to $t \in [1/4, 3/4]$, with

$$\Sigma = \begin{pmatrix} 1 & 1/(6\sqrt{3}) \\ 1/(6\sqrt{3}) & 1/36 \end{pmatrix}.$$

The response variable was then generated as

$$y_i = \begin{cases} 1, & \beta_0 x_i + \eta_0(t_i) + \varepsilon_i \geq 0, \\ 0, & \beta_0 x_i + \eta_0(t_i) + \varepsilon_i < 0, \end{cases}$$

where $\beta_0 = 2$, $\eta_0(t) = 2\sin(4\pi t)$ and $\varepsilon_i$ was a standard logistic variate. For each data set generated from this model, we also created three contaminated data sets, denoted $C_1$, $C_2$ and $C_3$ in Table 3. The purpose of the first two contaminations is to see how the robust methods work when one has contamination in $y$ only.

- CONTAMINATION 1. The contaminated data points were generated as follows: $u_i \sim \mathcal{U}(0,1)$, $x_i = x_i$ and

$$y_i = \begin{cases} y_i & \text{if } u_i \leq 0.90, \\ \text{a new observation from } Bi(1, 0.5) & \text{if } u_i > 0.90. \end{cases}$$

- CONTAMINATION 2. For each generated data set, we chose ten "design points" with $H(\beta_0 x_i + \eta_0(t_i)) > 0.99$, where $H(u) = 1/(1 + \exp(-u))$, so at those points, the conditional mean of $y$ given the covariates is not close to 0.5. We then contaminate $y$ as in Contamination 1, but only at those ten points. Of those ten points, about half are expected to be outliers with large Pearson residuals.
- CONTAMINATION 3. Here, we considered a contamination with bad leverage points by using $u_i \sim \mathcal{U}(0,1)$,

$$x_i = \begin{cases} x_i & \text{if } u_i \leq 0.90, \\ \text{a new observation from } N(10, 1) & \text{if } u_i > 0.90, \end{cases}$$

$$y_i = \begin{cases} y_i & \text{if } u_i \leq 0.90, \\ \text{a new observation from } Bi(1, 0.05) & \text{if } u_i > 0.90. \end{cases}$$

Both the original and the contaminated data sets were analyzed using the three competing estimators. Using a bandwidth of $h_n = 0.1$, we summarized the results in Table 3 based on 100 Monte Carlo samples. The bandwidth was chosen to be smaller than that used in Study 1 because we have 200 distinct observed values of $t$ here, as compared to ten in the earlier study. Table 3 shows the poor performance of the classical estimates of $\beta$, especially under contamination $C_3$. Under $C_1$, most contaminated $y$ do not result in large Pearson residuals and the robust estimators RQL and MOD can improve the nonrobust estimator somewhat, but not as significantly as under $C_2$ and $C_3$. With respect to the estimation of $\eta$, all procedures seem to be stable because the magnitude of outlying $y$ is very limited in this case.

Our studies show the good performance of the two families of robust estimators considered here in the presence of outliers. The MOD method often shows smaller bias for estimating $\boldsymbol{\beta}$, but its mean square error is usually similar to that of RQL.

ROBUST SEMIPARAMETRIC REGRESSION 15

# APPENDIX

## A.1. Proof of the consistency results.

PROOF OF THEOREM 3.1. (a) Let $Z_i(a,\boldsymbol{\beta}) = \rho(y_i, \mathbf{x}_i^{\mathrm{T}}\boldsymbol{\beta} + a)w_1(\mathbf{x}_i)$,

$$R_{1n}(a,\boldsymbol{\beta},t) = (nh_n)^{-1}\sum_{i=1}^{n} Z_i(a,\boldsymbol{\beta})K((t-t_i)/h_n),$$

$$R_{0n}(t) = (nh_n)^{-1}\sum_{i=1}^{n} K((t-t_i)/h_n).$$

Then $S_n(a,\boldsymbol{\beta},t) = R_{1n}(a,\boldsymbol{\beta},t)/R_{0n}(t)$, which implies that

$$\sup_{\substack{\boldsymbol{\beta}\in\mathcal{K} \\ a\in\mathbb{R}}} \|S_n(a,\boldsymbol{\beta},\cdot) - S(a,\boldsymbol{\beta},\cdot)\|_{0,\infty}$$

$$\leq \Big[\sup_{\substack{\boldsymbol{\beta}\in\mathcal{K} \\ a\in\mathbb{R}}} \|R_{1n}(a,\boldsymbol{\beta},\cdot) - E(R_{1n}(a,\boldsymbol{\beta},\cdot))\|_{0,\infty}$$

$$+ \sup_{\substack{\boldsymbol{\beta}\in\mathcal{K} \\ a\in\mathbb{R}}} \|E(R_{1n}(a,\boldsymbol{\beta},\cdot)) - S(a,\boldsymbol{\beta},\cdot)E(R_{0n}(\cdot))\|_{0,\infty}$$

$$+ \|\rho\|_\infty \|w_1\|_\infty \|R_{0n} - E(R_{0n})\|_{0,\infty}\Big]\Big\{\inf_{t\in\mathcal{T}_0} R_{0n}(t)\Big\}^{-1},$$

where $\|\rho\|_\infty = \sup_{(y,a)} |\rho(y,a)|$ and $\|w_1\|_\infty = \sup_{\mathbf{x}} |w_1(\mathbf{x})|$.

Since $E(R_{0n}(t)) = \int K(u) f_T(t-uh_n)\,du > A_1(\mathcal{T}_\delta)$, it is enough to show that

(A.1) $$\sup_{\substack{\boldsymbol{\beta}\in\mathcal{K} \\ a\in\mathbb{R}}} \|R_{1n}(a,\boldsymbol{\beta},\cdot) - E(R_{1n}(a,\boldsymbol{\beta},\cdot))\|_{0,\infty} \xrightarrow{a.s.} 0,$$

(A.2) $$\|R_{0n} - E(R_{0n})\|_{0,\infty} \xrightarrow{a.s.} 0,$$

(A.3) $$\sup_{\substack{\boldsymbol{\beta}\in\mathcal{K} \\ a\in\mathbb{R}}} \|E(R_{1n}(a,\boldsymbol{\beta},\cdot)) - S(a,\boldsymbol{\beta},\cdot)E(R_{0n}(\cdot))\|_{0,\infty} \to 0.$$

Assumptions C2–C4 imply (A.2); see [20], page 35. On the other hand, (A.3) follows easily from the boundness of $\rho$, the integrability of the kernel, the equicontinuity condition C5 and the fact that $h_n \to 0$. In order to prove (A.1), let us consider the class of functions

$$\mathcal{F}_n = \{f_{t,a,\boldsymbol{\beta},h_n}(y,\mathbf{x},v) = B^{-1}\rho(y,\mathbf{x}^{\mathrm{T}}\boldsymbol{\beta}+a)w_1(\mathbf{x})K_{t,h_n}(v)\},$$

with $B = \|\rho\|_\infty \|w_1\|_\infty \|K\|_\infty$ and $K_{t,h_n}(v) = K((t-v)/h_n)$. Using the fact that the graphs of translated kernels $K_{t,h_n}$ have polynomial discrimination, inequality $0 \leq K_{t,h_n} \leq \|K\|_\infty$ and assumption (ii), we obtain that $N(\varepsilon, \mathcal{F}_n, L^1(\mathbb{Q})) \leq A_1 \varepsilon^{-W_1}$ for any probability $\mathbb{Q}$ and $0 < \varepsilon < 1$, where $A_1$ and $W_1$ do not depend on $n$. Since for any $f_{t,a,\boldsymbol{\beta},h_n} \in \mathcal{F}_n$, $|f_{t,a,\boldsymbol{\beta},h}| \leq 1$ and $E(f_{t,a,\boldsymbol{\beta},h_n}^2(y,\mathbf{x},v)) \leq h_n \|K\|_\infty^{-1} \|f_T\|_\infty$, Theorem 37 in [20] and C4 imply that

$$(h_n)^{-1} \sup_{\mathcal{F}_n} \left| n^{-1} \sum_{i=1}^n f_{t,a,\boldsymbol{\beta},h_n}(y_i, \mathbf{x}_i, t_i) - E f_{t,a,\boldsymbol{\beta},h_n}(y_1, \mathbf{x}_1, t_1) \right| \xrightarrow{a.s.} 0,$$

which concludes the proof of (A.1).

(b) The continuity of $\eta_{\boldsymbol{\beta}}(t)$ implies that $\eta_{\boldsymbol{\beta}}(t)$ is bounded for $t \in \mathcal{T}_0$ and $\boldsymbol{\beta} \in \mathcal{K}$ and, thus, that there exists a compact set $\mathcal{A}(\mathcal{T}_0, \mathcal{K})$ such that $\eta_{\boldsymbol{\beta}}(t) \in \mathcal{A}(\mathcal{T}_0, \mathcal{K})$ for any $t \in \mathcal{T}_0$ and $\boldsymbol{\beta} \in \mathcal{K}$. Assume that $\sup_{\boldsymbol{\beta} \in \mathcal{K}} \|\hat{\eta}_{\boldsymbol{\beta}} - \eta_{\boldsymbol{\beta}}\|_{0,\infty}$ does not converge to 0 in a set $\Omega_0$ with $P(\Omega_0) > 0$. Then for each $\omega \in \Omega_0$, we have that there exists a sequence $(\boldsymbol{\beta}_k, t_k)$ such that $t_k \in \mathcal{T}_0$, $\boldsymbol{\beta}_k \in \mathcal{K}$ and $\hat{\eta}_{\boldsymbol{\beta}_k}(t_k) - \eta_{\boldsymbol{\beta}_k}(t_k) \to c \neq 0$. Since $\mathcal{T}_0$ and $\mathcal{K}$ are compact, without loss of generality we can assume that $t_k \to t_L \in \mathcal{T}_0$ and $\boldsymbol{\beta}_k \to \boldsymbol{\beta}_L \in \mathcal{K}$ and hence obtain that $\eta_{\boldsymbol{\beta}_k}(t_k) \to \eta_{\boldsymbol{\beta}_L}(t_L)$, implying that $\hat{\eta}_{\boldsymbol{\beta}_k}(t_k) - \eta_{\boldsymbol{\beta}_L}(t_L) \to c$. When $c < \infty$, the same steps as those used in Lemma A1 of [11] lead to a contradiction. If $c = \infty$, we have that $\hat{\eta}_{\boldsymbol{\beta}_k}(t_k) \to \infty$. By assumption, we have that

$$0 < i = \inf_{\substack{\boldsymbol{\beta} \in \mathcal{K} \\ t \in \mathcal{T}_0}} \left[ \lim_{|a| \to \infty} S(a, \boldsymbol{\beta}, t) - S(\eta_{\boldsymbol{\beta}}(t), \boldsymbol{\beta}, t) \right]$$

and so $\lim_{|a| \to \infty} S(a, \boldsymbol{\beta}_L, t_L) - S(\eta_{\boldsymbol{\beta}_L}(t_L), \boldsymbol{\beta}_L, t_L) \geq i$. Thus, for $k$ sufficiently large, $S(\hat{\eta}_{\boldsymbol{\beta}_k}(t_k), \boldsymbol{\beta}_L, t_L) > S(\eta_{\boldsymbol{\beta}_L}(t_L), \boldsymbol{\beta}_L, t_L) + i/2$. The equicontinuity condition implies that given $\varepsilon > 0$, for $k$ sufficiently large, $S(\eta_{\boldsymbol{\beta}_L}(t_L), \boldsymbol{\beta}_k, t_k) \leq S(\eta_{\boldsymbol{\beta}_L}(t_L), \boldsymbol{\beta}_L, t_L) + \varepsilon/4$ and $S(\hat{\eta}_{\boldsymbol{\beta}_k}(t_k), \boldsymbol{\beta}_L, t_L) \leq S(\hat{\eta}_{\boldsymbol{\beta}_k}(t_k), \boldsymbol{\beta}_k, t_k) + \varepsilon/4$, which from (a) and the definition of $\hat{\eta}_{\boldsymbol{\beta}}$, implies that $S(\hat{\eta}_{\boldsymbol{\beta}_k}(t_k), \boldsymbol{\beta}_L, t_L) \leq S_n(\hat{\eta}_{\boldsymbol{\beta}_k}(t_k), \boldsymbol{\beta}_k, t_k) + \varepsilon/2 \leq S_n(\eta_{\boldsymbol{\beta}_L}(t_L), \boldsymbol{\beta}_k, t_k) + \varepsilon/2$. Again using (a), we obtain $S(\hat{\eta}_{\boldsymbol{\beta}_k}(t_k), \boldsymbol{\beta}_L, t_L) \leq S_n(\eta_{\boldsymbol{\beta}_L}(t_L), \boldsymbol{\beta}_k, t_k) + \varepsilon/2 \leq S(\eta_{\boldsymbol{\beta}_L}(t_L), \boldsymbol{\beta}_k, t_k) + 3\varepsilon/4 \leq S(\eta_{\boldsymbol{\beta}_L}(t_L), \boldsymbol{\beta}_L, t_L) + \varepsilon$. Hence, for $k$ sufficiently large, $S(\eta_{\boldsymbol{\beta}_L}(t_L), \boldsymbol{\beta}_L, t_L) + i/2 < S(\hat{\eta}_{\boldsymbol{\beta}_k}(t_k), \boldsymbol{\beta}_L, t_L) \leq S(\eta_{\boldsymbol{\beta}_L}(t_L), \boldsymbol{\beta}_L, t_L) + \varepsilon$, leading to a contradiction. $\square$

The next proposition states a general uniform convergence result which will be helpful in proving Theorems 3.2 and 4.1.

We will begin by fixing some notation. Denote by $\mathcal{C}^1(\mathcal{T})$ the set of continuously differentiable functions in $\mathcal{T}$. Note that if $S^1(a, \boldsymbol{\beta}, \tau)$ defined in (12) is continuously differentiable with respect to $(a, \tau)$, then $\eta_{\boldsymbol{\beta}} \in \mathcal{C}^1(\mathcal{T})$. $\mathcal{V}(\boldsymbol{\beta})$ and $\mathcal{H}_\delta(\boldsymbol{\beta})$ denote neighborhoods of $\boldsymbol{\beta} \in \mathcal{K}$ and $\eta_{\boldsymbol{\beta}}$ such that $\mathcal{V}(\boldsymbol{\beta}) \subset \mathcal{K}$ and

$$\mathcal{H}_\delta(\boldsymbol{\beta}) = \left\{ u \in \mathcal{C}^1(\mathcal{T}) : \|u - \eta_{\boldsymbol{\beta}}\|_\infty \leq \delta, \left\| \frac{\partial}{\partial t} u - \frac{\partial}{\partial t} \eta_{\boldsymbol{\beta}} \right\|_\infty \leq \delta \right\}.$$

PROPOSITION A.1. *Let $(y_i, \mathbf{x}_i, t_i)$ be independent observations such that $y_i|(\mathbf{x}_i, t_i) \sim F(\cdot, \mu_i)$, with $\mu_i = H(\eta_0(t_i) + \mathbf{x}_i^{\mathrm{T}} \boldsymbol{\beta}_0)$ and $\mathrm{VAR}\,(y_i|(\mathbf{x}_i, t_i)) = V(\mu_i)$. Assume that $t_i$ are random variables with distribution on $\mathcal{T}$. Let $g: \mathbb{R}^2 \to \mathbb{R}$ be a continuous and bounded function, $W(\mathbf{x}, t): \mathbb{R}^{p+1} \to \mathbb{R}$ be such that $E(|W(\mathbf{x}, t)|) < \infty$ and $\eta_{\boldsymbol{\beta}}(t) = \eta(\boldsymbol{\beta}, t): \mathbb{R}^{p+1} \to \mathbb{R}$ be a continuous function of $(\boldsymbol{\beta}, t)$. Define $L(y, \mathbf{x}, t, \boldsymbol{\beta}, v) = g(y, \mathbf{x}^{\mathrm{T}} \boldsymbol{\beta} + v(t)) W(\mathbf{x}, t)$ and $\mathbf{E}(\boldsymbol{\beta}) = E_0(L(y, \mathbf{x}, t, \boldsymbol{\beta}, \eta_{\boldsymbol{\beta}}))$. Then*

(a) $E(n^{-1} \sum_{i=1}^n L(y_i, \mathbf{x}_i, t_i, \boldsymbol{\theta}, v)) \to \mathbf{E}(\boldsymbol{\beta})$ when $\|\boldsymbol{\theta} - \boldsymbol{\beta}\| + \|v - \eta_{\boldsymbol{\beta}}\|_\infty \to 0$,

(b) $\sup_{\boldsymbol{\theta} \in \mathcal{K}} |n^{-1} \sum_{i=1}^n L(y_i, \mathbf{x}_i, t_i, \boldsymbol{\theta}, \eta_{\boldsymbol{\theta}}) - E(L(y_i, \mathbf{x}_i, t_i, \boldsymbol{\theta}, \eta_{\boldsymbol{\theta}}))| \xrightarrow{a.s.} 0$,

(c) $\sup_{\boldsymbol{\theta} \in \mathcal{K}, v \in \mathcal{H}_1(\boldsymbol{\beta})} |n^{-1} \sum_{i=1}^n L(y_i, \mathbf{x}_i, t_i, \boldsymbol{\theta}, v) - E(L(y_i, \mathbf{x}_i, t_i, \boldsymbol{\theta}, v))| \xrightarrow{a.s.} 0$

*if, in addition, $\mathcal{T}$ is compact and $\eta_{\boldsymbol{\beta}} \in \mathcal{C}^1(\mathcal{T})$.*

PROOF. (a) follows from the dominated convergence theorem. The proofs of (b) and (c) follow using the continuity of $\eta_{\boldsymbol{\beta}}$ and $g$, Theorem 3 in Chapter 2 of [20], the compactness of $\mathcal{K}$ and $\mathcal{H}_1(\boldsymbol{\beta})$ and analogous arguments to those considered in [2]. $\square$

REMARK A.1. Proposition A.1 implies that for any weakly consistent estimate $\hat{\eta}_{\boldsymbol{\beta}}$ of $\eta_{\boldsymbol{\beta}}$ such that $\sup_{t \in \mathcal{T}} |(\partial/\partial t) \hat{\eta}_{\boldsymbol{\beta}}(t) - (\partial/\partial t) \eta_{\boldsymbol{\beta}}(t)| \xrightarrow{a.s.} 0$ and $\sup_{t \in \mathcal{T}} |\hat{\eta}_{\boldsymbol{\beta}}(t) - \eta_{\boldsymbol{\beta}}(t)| \xrightarrow{a.s.} 0$, we have $(1/n) \sum_{i=1}^n \mathbf{H}(y_i, \mathbf{x}_i, t_i, \boldsymbol{\beta}, \hat{\eta}_{\boldsymbol{\beta}}) \xrightarrow{a.s.} \mathbf{E}(\boldsymbol{\beta})$. An analogous result can be obtained replacing $\xrightarrow{a.s.}$ by $\xrightarrow{p}$.

PROOF OF THEOREM 3.2. (a) Define

$$\tilde{F}_n(\boldsymbol{\beta}) = \frac{1}{n} \sum_{i=1}^n \rho(y_i, \mathbf{x}_i^{\mathrm{T}} \boldsymbol{\beta} + \eta_{\boldsymbol{\beta}}(t_i)) w_2(\mathbf{x}_i).$$

For any $\varepsilon > 0$, let $\mathcal{T}_0$ be a compact set such that $P(t_i \notin \mathcal{T}_0) < \varepsilon$ and let $T_n = (1/n) \sum_{i=1}^n I(t_i \notin \mathcal{T}_0)$. We then have

$$\sup_{\boldsymbol{\beta} \in \mathcal{K}} |F_n(\boldsymbol{\beta}) - \tilde{F}_n(\boldsymbol{\beta})| \leq \|w_2\|_\infty \Big\{ \|\Psi\|_\infty \sup_{\boldsymbol{\beta} \in \mathcal{K}} \|\hat{\eta}_{\boldsymbol{\beta}} - \eta_{\boldsymbol{\beta}}\|_{0,\infty} + 2\|\rho\|_\infty T_n \Big\}.$$

Hence, using (16) and the strong law of large numbers, we easily get that

(A.4) $$\sup_{\boldsymbol{\beta} \in \mathcal{K}} |F_n(\boldsymbol{\beta}) - \tilde{F}_n(\boldsymbol{\beta})| \xrightarrow{a.s.} 0.$$

Moreover, Proposition A.1(b) with $W(\mathbf{x}, t) = w_2(\mathbf{x})$ and $g(y, u) = \rho(y, u)$ implies that $\sup_{\boldsymbol{\beta} \in \mathcal{K}} |\tilde{F}_n(\boldsymbol{\beta}) - F(\boldsymbol{\beta})| \xrightarrow{a.s.} 0$ which, together with (A.4), concludes the proof of (a).

(b) Note that (a) implies that $F_n(\hat{\boldsymbol{\beta}}) = \inf_{\boldsymbol{\beta} \in \mathcal{K}} F_n(\boldsymbol{\beta}) \xrightarrow{a.s.} \inf_{\boldsymbol{\beta} \in \mathcal{K}} F(\boldsymbol{\beta}) = F(\boldsymbol{\beta}_0)$ and $F_n(\hat{\boldsymbol{\beta}}) - F(\hat{\boldsymbol{\beta}}) \xrightarrow{a.s.} 0$, and so $F(\hat{\boldsymbol{\beta}}) \xrightarrow{a.s.} F(\boldsymbol{\beta}_0)$. Since $F$ has a unique minimum at $\boldsymbol{\beta}_0$, (b) follows easily. $\square$

**A.2. Proof of the asymptotic normality of the regression estimates.** For the sake of simplicity, we denote

$$\chi(y,a) = \frac{\partial}{\partial a}\Psi(y,a), \qquad \chi_1(y,a) = \frac{\partial^2}{\partial a^2}\Psi(y,a),$$

(A.5) $$\hat{v}(\boldsymbol{\beta},t) = \hat{\eta}_{\boldsymbol{\beta}}(t) - \eta_{\boldsymbol{\beta}}(t), \qquad \hat{v}_0(t) = \hat{v}(\boldsymbol{\beta}_0,t),$$

(A.6) $$\hat{v}_j(\boldsymbol{\beta},t) = \frac{\partial \hat{v}(\boldsymbol{\beta},t)}{\partial \beta_j}, \qquad \hat{v}_{j,0}(t) = \hat{v}_j(\boldsymbol{\beta}_0,t).$$

We list a set of conditions needed for the asymptotic normality theorem, followed by general comments on those conditions. The first condition is on the preliminary estimate of $\eta_{\boldsymbol{\beta}}(t)$ and the rest are on the score functions and the underlying model distributions.

N1. (a) The functions $\hat{\eta}_{\boldsymbol{\beta}}(t)$ and $\eta_{\boldsymbol{\beta}}(t)$ are continuously differentiable with respect to $(\boldsymbol{\beta},t)$ and twice continuously differentiable with respect to $\boldsymbol{\beta}$ such that $(\partial^2 \eta_{\boldsymbol{\beta}}(t))/\partial \beta_j \partial \beta_\ell|_{\boldsymbol{\beta}=\boldsymbol{\beta}_0}$ is bounded. Furthermore, for any $1 \leq j, \ell \leq p$, $(\partial^2 \eta_{\boldsymbol{\beta}}(t))/\partial \beta_j \partial \beta_\ell$ satisfies the following equicontinuity condition:

$$\forall \varepsilon > 0, \; \exists \delta > 0 : |\boldsymbol{\beta}_1 - \boldsymbol{\beta}_0| < \delta$$

$$\Rightarrow \left\| \frac{\partial^2}{\partial \beta_j \partial \beta_\ell} \eta_{\boldsymbol{\beta}} \bigg|_{\boldsymbol{\beta}=\boldsymbol{\beta}_1} - \frac{\partial^2}{\partial \beta_j \partial \beta_\ell} \eta_{\boldsymbol{\beta}} \bigg|_{\boldsymbol{\beta}=\boldsymbol{\beta}_0} \right\|_\infty < \varepsilon.$$

(b) $\|\hat{\eta}_{\hat{\boldsymbol{\beta}}} - \eta_0\|_\infty \xrightarrow{p} 0$ for any consistent estimate $\hat{\boldsymbol{\beta}}$ of $\boldsymbol{\beta}_0$.

(c) For each $t \in \mathcal{T}$ and $\boldsymbol{\beta}$, $\hat{v}(\boldsymbol{\beta},t) \xrightarrow{p} 0$. Moreover, $n^{1/4}\|\hat{v}_0\|_\infty \xrightarrow{p} 0$ and $n^{1/4}\|\hat{v}_{j,0}\|_\infty \xrightarrow{p} 0$ for all $1 \leq j \leq p$.

(d) There exists a neighborhood of $\boldsymbol{\beta}_0$ with closure $\mathcal{K}$ such that for any $1 \leq j, \ell \leq p$, $\sup_{\boldsymbol{\beta} \in \mathcal{K}}(\|\hat{v}_j(\boldsymbol{\beta},\cdot)\|_\infty + \|(\partial \hat{v}_j(\boldsymbol{\beta},\cdot))/\partial \beta_\ell\|_\infty) \xrightarrow{p} 0$.

(e) $\|(\partial \hat{v}_0)/\partial t\|_\infty + \|(\partial \hat{v}_{j,0})/\partial t\|_\infty \xrightarrow{p} 0$ for any $1 \leq j \leq p$.

N2. The functions $\Psi$, $\chi$, $\chi_1$, $w_2$ and $\psi_2(\mathbf{x}) = \mathbf{x}w_2(\mathbf{x})$ are bounded and continuous.

N3. The matrix $\mathbf{A}$ is nonsingular, where

$$\mathbf{A} = \mathrm{E}_0\left[\left\{\chi(y,\mathbf{x}^{\mathrm{T}}\boldsymbol{\beta}_0 + \eta_0(t))\left[\mathbf{x} + \frac{\partial}{\partial \boldsymbol{\beta}}\eta_{\boldsymbol{\beta}}(t)\bigg|_{\boldsymbol{\beta}=\boldsymbol{\beta}_0}\right]\left[\mathbf{x} + \frac{\partial}{\partial \boldsymbol{\beta}}\eta_{\boldsymbol{\beta}}(t)\bigg|_{\boldsymbol{\beta}=\boldsymbol{\beta}_0}\right]^{\mathrm{T}}\right.\right.$$
$$\left.\left. + \Psi(y,\mathbf{x}^{\mathrm{T}}\boldsymbol{\beta}_0 + \eta_0(t))\frac{\partial^2}{\partial \boldsymbol{\beta}\partial \boldsymbol{\beta}^{\mathrm{T}}}\eta_{\boldsymbol{\beta}}(t)\bigg|_{\boldsymbol{\beta}=\boldsymbol{\beta}_0}^{\mathrm{T}}\right\}w_2(\mathbf{x})\right].$$

N4. The matrix $\boldsymbol{\Sigma}$ is positive definite with

$$\boldsymbol{\Sigma} = \mathrm{E}_0\left\{\Psi^2(y,\mathbf{x}^{\mathrm{T}}\boldsymbol{\beta}_0 + \eta_0(t))w_2^2(\mathbf{x})\left[\mathbf{x} + \frac{\partial}{\partial \boldsymbol{\beta}}\eta_{\boldsymbol{\beta}}(t)\bigg|_{\boldsymbol{\beta}=\boldsymbol{\beta}_0}\right]\right.$$

$$\times \left[\mathbf{x} + \frac{\partial}{\partial \boldsymbol{\beta}} \eta_{\boldsymbol{\beta}}(t)\Big|_{\boldsymbol{\beta}=\boldsymbol{\beta}_0}\right]^{\mathrm{T}}\bigg\}.$$

N5. (a) $\mathrm{E}_0\{\Psi(y, \mathbf{x}^{\mathrm{T}}\boldsymbol{\beta}_0 + \eta_0(t))|(\mathbf{x}, t)\} = 0$.

(b) $\mathrm{E}_0[\{\chi(y, \mathbf{x}^{\mathrm{T}}\boldsymbol{\beta}_0 + \eta_0(\tau))(\mathbf{x} + (\partial \eta_{\boldsymbol{\beta}}(\tau))/\partial \boldsymbol{\beta}|_{\boldsymbol{\beta}=\boldsymbol{\beta}_0})\}w_2(\mathbf{x})|t=\tau] = 0$.

N6. $\mathrm{E}_0(w_2(\mathbf{x})\|\mathbf{x} + (\partial \eta_{\boldsymbol{\beta}}(\tau))/\partial \boldsymbol{\beta}|_{\boldsymbol{\beta}=\boldsymbol{\beta}_0}\|^2) < \infty$.

REMARK A.2. Conditions N1(a) and (d) imply that for any consistent estimator $\tilde{\boldsymbol{\beta}}$ of $\boldsymbol{\beta}_0$, we have $\Delta_n \xrightarrow{p} 0$ and $\Lambda_n \xrightarrow{p} 0$ with

$$\Delta_n = \max_{1 \leq j \leq p} \left\|\frac{\partial}{\partial \beta_j}\hat{\eta}_{\boldsymbol{\beta}}\Big|_{\boldsymbol{\beta}=\tilde{\boldsymbol{\beta}}} - \frac{\partial}{\partial \beta_j}\eta_{\boldsymbol{\beta}}\Big|_{\boldsymbol{\beta}=\boldsymbol{\beta}_0}\right\|_\infty,$$

$$\Lambda_n = \max_{1 \leq j,\ell \leq p} \left\|\frac{\partial^2}{\partial \beta_j \partial \beta_\ell}\hat{\eta}_{\boldsymbol{\beta}}\Big|_{\boldsymbol{\beta}=\tilde{\boldsymbol{\beta}}} - \frac{\partial^2}{\partial \beta_j \partial \beta_\ell}\eta_{\boldsymbol{\beta}}\Big|_{\boldsymbol{\beta}=\boldsymbol{\beta}_0}\right\|_\infty.$$

Condition N1(b) follows from the continuity of $\eta_{\boldsymbol{\beta}}(t) = \eta(\boldsymbol{\beta}, t)$ with respect to $(\boldsymbol{\beta}, t)$ and Theorem 3.1 that leads to $\sup_{\boldsymbol{\beta} \in \mathcal{K}} \|\hat{\eta}_{\boldsymbol{\beta}} - \eta_{\boldsymbol{\beta}}\|_\infty \xrightarrow{a.s.} 0$.

REMARK A.3. When the kernel $K$ is continuously differentiable with derivative $K'$ bounded and with bounded variation, the uniform convergence required in N1(b)–(e) can be derived using arguments analogous to those considered in Theorem 3.1 by using the facts that

$$\frac{\partial}{\partial t}\hat{\eta}_{\boldsymbol{\beta}}(t) = -\frac{(nh_n^2)^{-1}\sum_{i=1}^n K'((t-t_i)/h_n)\Psi(y_i, \mathbf{x}_i^{\mathrm{T}}\boldsymbol{\beta} + \hat{\eta}_{\boldsymbol{\beta}}(t))}{(nh_n)^{-1}\sum_{i=1}^n K((t-t_i)/h_n)\chi(y_i, \mathbf{x}_i^{\mathrm{T}}\boldsymbol{\beta} + \hat{\eta}_{\boldsymbol{\beta}}(t))},$$

$$\frac{\partial}{\partial \beta_j}\hat{\eta}_{\boldsymbol{\beta}}(t) = -\frac{(nh_n)^{-1}\sum_{i=1}^n K((t-t_i)/h_n)\chi(y_i, \mathbf{x}_i^{\mathrm{T}}\boldsymbol{\beta} + \hat{\eta}_{\boldsymbol{\beta}}(t))x_{ij}}{(nh_n)^{-1}\sum_{i=1}^n K((t-t_i)/h_n)\chi(y_i, \mathbf{x}_i^{\mathrm{T}}\boldsymbol{\beta} + \hat{\eta}_{\boldsymbol{\beta}}(t))}$$

and requiring that

$$\sup_{t \in \mathcal{T}} E\Big(\sup_{\boldsymbol{\beta} \in \mathcal{K}} |\chi(y_1, \mathbf{x}_1^{\mathrm{T}}\boldsymbol{\beta} + \eta_{\boldsymbol{\beta}}(t))|\|\mathbf{x}_1\| \mid t_1 = t\Big) < \infty,$$

$$\sup_{t \in \mathcal{T}} E\Big(\sup_{\boldsymbol{\beta} \in \mathcal{K}} |\chi_1(y_1, \mathbf{x}_1^{\mathrm{T}}\boldsymbol{\beta} + \eta_{\boldsymbol{\beta}}(t))|\|\mathbf{x}_1\| \mid t_1 = t\Big) < \infty,$$

$$\inf_{\substack{\boldsymbol{\beta} \in \mathcal{K} \\ t \in \mathcal{T}}} E(\chi(y_1, \mathbf{x}_1^{\mathrm{T}}\boldsymbol{\beta} + \eta_{\boldsymbol{\beta}}(t)) \mid t_1 = t) > 0.$$

The uniform convergence rates required in N1(c) are fulfilled when $\hat{\eta}_{\boldsymbol{\beta}}$ is defined as in (7) and a rate-optimal bandwidth is used for the kernel. The convergence requirements in N1 are analogous to those required in condition (7) in [21], page 510 and are needed in order to obtain the desired rate of convergence for the regression estimates. More precisely, assumption N1(c) avoids the bias term and ensures that $G_n(\hat{\eta}_{\boldsymbol{\beta}_0})$ will behave asymptotically as $G_n(\eta_{\boldsymbol{\beta}_0})$, where for any $\boldsymbol{\beta} \in \mathbb{R}^p$ and any differentiable function $v_{\boldsymbol{\beta}}(t) = v(\boldsymbol{\beta},t) : \mathbb{R}^{p+1} \to \mathbb{R}$,

$$G_n(v_{\boldsymbol{\beta}}) = \frac{1}{\sqrt{n}} \sum_{i=1}^n \Psi(y_i, \mathbf{x}_i^{\mathrm{T}}\boldsymbol{\beta}_0 + v_{\boldsymbol{\beta}_0}(t_i))\left[\mathbf{x}_i + \frac{\partial}{\partial \boldsymbol{\beta}}v_{\boldsymbol{\beta}}(t_i)\bigg|_{\boldsymbol{\beta}=\boldsymbol{\beta}_0}\right]w_2(\mathbf{x}_i).$$

REMARK A.4. If N4 is fulfilled, then the columns of $\mathbf{x} + (\partial \eta_{\boldsymbol{\beta}}(t))/\partial \boldsymbol{\beta}|_{\boldsymbol{\beta}=\boldsymbol{\beta}_0}$ will not be collinear. It is necessary not to allow $\mathbf{x}$ to be predicted by $t$ to get root-$n$ regression estimates.

Note that for the $\Psi$ functions considered by Bianco and Yohai [5], Croux and Haesbroeck [12] and Cantoni and Ronchetti [9], N5(a) is satisfied. This condition is the conditional Fisher consistency property as stated by Künsch, Stefanski and Carroll [17] for the generalized linear regression model.

Note also that N5(b) is fulfilled if $w_2 \equiv w_1$. Effectively, since $\eta_{\boldsymbol{\beta}}(\tau)$ minimizes $S(a,\boldsymbol{\beta},\tau)$ for each $\tau$, it satisfies

$$E_0[\Psi(y, \mathbf{x}^{\mathrm{T}}\boldsymbol{\beta} + \eta_{\boldsymbol{\beta}}(\tau))w_1(\mathbf{x}) \mid t = \tau] = 0,$$

thus, differentiating with respect to $\boldsymbol{\beta}$, we get

$$E_0\left[\chi(y, \mathbf{x}^{\mathrm{T}}\boldsymbol{\beta} + \eta_{\boldsymbol{\beta}}(\tau))\left(\mathbf{x} + \frac{\partial}{\partial \boldsymbol{\beta}}\eta_{\boldsymbol{\beta}}(\tau)\right)w_1(\mathbf{x}) \mid t = \tau\right] = 0.$$

Moreover, if either $w_2 \equiv w_1$ or N5(a) holds, then

$$\mathbf{A} = E_0\left\{\chi(y, \mathbf{x}^{\mathrm{T}}\boldsymbol{\beta}_0 + \eta_0(t))\left[\mathbf{x} + \frac{\partial}{\partial \boldsymbol{\beta}}\eta_{\boldsymbol{\beta}}(t)\bigg|_{\boldsymbol{\beta}=\boldsymbol{\beta}_0}\right]\left[\mathbf{x} + \frac{\partial}{\partial \boldsymbol{\beta}}\eta_{\boldsymbol{\beta}}(t)\bigg|_{\boldsymbol{\beta}=\boldsymbol{\beta}_0}\right]^{\mathrm{T}}w_2(\mathbf{x})\right\}.$$

Therefore, if $\Psi(y,u)$ is strictly monotone in $u$ and $P(w_2(\mathbf{x}) > 0) = 1$, then N3 holds, that is, $\mathbf{A}$ will be nonsingular unless $P(\mathbf{a}^{\mathrm{T}}[\mathbf{x} + (\partial \eta_{\boldsymbol{\beta}}(t))/\partial \boldsymbol{\beta}|_{\boldsymbol{\beta}=\boldsymbol{\beta}_0}] = 0) = 1$ for some $\mathbf{a} \in \mathbb{R}^p$ (i.e., unless there is a linear combination of $\mathbf{x}$ which can be completely determined by $t$).

Assumption N6 is used to ensure the consistency of the estimates of $\mathbf{A}$ based on preliminary estimates of the regression parameter $\boldsymbol{\beta}$ and of the functions $\eta_{\boldsymbol{\beta}}$.

LEMMA A.1. *Let $(y_i, \mathbf{x}_i, t_i)$ be independent observations such that $y_i|(\mathbf{x}_i, t_i) \sim F(\cdot, \mu_i)$ with $\mu_i = H(\eta_0(t_i) + \mathbf{x}_i^{\mathrm{T}}\boldsymbol{\beta}_0)$ and $\mathrm{VAR}(y_i|(\mathbf{x}_i, t_i)) = V(\mu_i)$. Assume that $t_i$ are random variables with distribution on a compact set $\mathcal{T}$ and that N1–N3 and N6 hold. Let $\tilde{\boldsymbol{\beta}}$ be such that $\tilde{\boldsymbol{\beta}} \xrightarrow{p} \boldsymbol{\beta}$. Then $\mathbf{A}_n \xrightarrow{p} \mathbf{A}$, where $\mathbf{A}$ is given in N3, $\hat{\mathbf{z}}_i(\tilde{\boldsymbol{\beta}}) = \mathbf{x}_i + (\partial \hat{\eta}_{\boldsymbol{\beta}}(t_i))/\partial \boldsymbol{\beta}|_{\boldsymbol{\beta}=\tilde{\boldsymbol{\beta}}}$ and*

$$\mathbf{A}_n = n^{-1} \sum_{i=1}^n \chi(y_i, \mathbf{x}_i^{\mathrm{T}}\tilde{\boldsymbol{\beta}} + \hat{\eta}_{\tilde{\boldsymbol{\beta}}}(t_i))\hat{\mathbf{z}}_i(\tilde{\boldsymbol{\beta}})\hat{\mathbf{z}}_i(\tilde{\boldsymbol{\beta}})^{\mathrm{T}}w_2(\mathbf{x}_i)$$

$$+ n^{-1}\sum_{i=1}^{n} \Psi(y_i, \mathbf{x}_i^{\mathrm{T}}\tilde{\boldsymbol{\beta}} + \hat{\eta}_{\tilde{\boldsymbol{\beta}}}(t_i))\frac{\partial^2}{\partial\boldsymbol{\beta}\partial\boldsymbol{\beta}^{\mathrm{T}}}\hat{\eta}_{\boldsymbol{\beta}}(t_i)\bigg|_{\boldsymbol{\beta}=\tilde{\boldsymbol{\beta}}}^{\mathrm{T}} w_2(\mathbf{x}_i).$$

PROOF. The proof follows easily using a Taylor expansion, the required assumptions, Proposition A.1 and the fact that $\tilde{\boldsymbol{\beta}} \xrightarrow{p} \boldsymbol{\beta}_0$. Details can be found in [8]. □

PROOF OF THEOREM 4.1. Let $\hat{\boldsymbol{\beta}}$ be a solution of $F_n^1(\boldsymbol{\beta}) = 0$ defined in (15). Using a Taylor expansion of order one, we get

$$0 = \sum_{i=1}^{n} \Psi(y_i, \mathbf{x}_i^{\mathrm{T}}\boldsymbol{\beta}_0 + \hat{\eta}_{\boldsymbol{\beta}_0}(t_i))w_2(\mathbf{x}_i)\bigg[\mathbf{x}_i + \frac{\partial}{\partial\boldsymbol{\beta}}\hat{\eta}_{\boldsymbol{\beta}}(t_i)\bigg|_{\boldsymbol{\beta}=\boldsymbol{\beta}_0}\bigg] + n\mathbf{A}_n(\hat{\boldsymbol{\beta}} - \boldsymbol{\beta}_0),$$

where

$$\mathbf{A}_n = n^{-1}\sum_{i=1}^{n} \frac{\partial}{\partial\boldsymbol{\beta}}\bigg\{\Psi(y_i, \mathbf{x}_i^{\mathrm{T}}\boldsymbol{\beta} + \hat{\eta}_{\boldsymbol{\beta}}(t_i))\bigg[\mathbf{x}_i + \frac{\partial}{\partial\boldsymbol{\beta}}\hat{\eta}_{\boldsymbol{\beta}}(t_i)\bigg]\bigg\}\bigg|_{\boldsymbol{\beta}=\tilde{\boldsymbol{\beta}}} w_2(\mathbf{x}_i)$$

$$= n^{-1}\sum_{i=1}^{n} \chi(y_i, \mathbf{x}_i^{\mathrm{T}}\tilde{\boldsymbol{\beta}} + \hat{\eta}_{\tilde{\boldsymbol{\beta}}}(t_i))\bigg[\mathbf{x}_i + \frac{\partial}{\partial\boldsymbol{\beta}}\hat{\eta}_{\boldsymbol{\beta}}(t_i)\bigg|_{\boldsymbol{\beta}=\tilde{\boldsymbol{\beta}}}\bigg]\bigg[\mathbf{x}_i + \frac{\partial}{\partial\boldsymbol{\beta}}\hat{\eta}_{\boldsymbol{\beta}}(t_i)\bigg|_{\boldsymbol{\beta}=\tilde{\boldsymbol{\beta}}}\bigg]^{\mathrm{T}}$$

$$\times w_2(\mathbf{x}_i) + n^{-1}\sum_{i=1}^{n} \Psi(y_i, \mathbf{x}_i^{\mathrm{T}}\tilde{\boldsymbol{\beta}} + \hat{\eta}_{\tilde{\boldsymbol{\beta}}}(t_i))\frac{\partial^2}{\partial\boldsymbol{\beta}\partial\boldsymbol{\beta}^{\mathrm{T}}}\hat{\eta}_{\boldsymbol{\beta}}(t_i)\bigg|_{\boldsymbol{\beta}=\tilde{\boldsymbol{\beta}}}^{\mathrm{T}} w_2(\mathbf{x}_i)$$

with $\tilde{\boldsymbol{\beta}}$ an intermediate point between $\boldsymbol{\beta}$ and $\hat{\boldsymbol{\beta}}$. Note that in the partially linear regression model, only the first term in $\mathbf{A}_n$ is different from 0 since $\hat{\eta}_{\boldsymbol{\beta}}(t)$ is linear in $\boldsymbol{\beta}$.

From Lemma A.1, we have that $\mathbf{A}_n \xrightarrow{p} \mathbf{A}$, where $\mathbf{A}$ is defined in N3. Therefore, in order to obtain the asymptotic distribution of $\hat{\boldsymbol{\beta}}$, it will be enough to derive the asymptotic behavior of

$$\hat{L}_n = n^{-1/2}\sum_{i=1}^{n} \Psi(y_i, \mathbf{x}_i^{\mathrm{T}}\boldsymbol{\beta}_0 + \hat{\eta}_{\boldsymbol{\beta}_0}(t_i))\bigg[\mathbf{x}_i + \frac{\partial}{\partial\boldsymbol{\beta}}\hat{\eta}_{\boldsymbol{\beta}}(t_i)\bigg|_{\boldsymbol{\beta}=\boldsymbol{\beta}_0}\bigg] w_2(\mathbf{x}_i).$$

Let

$$L_n = n^{-1/2}\sum_{i=1}^{n} \Psi(y_i, \mathbf{x}_i^{\mathrm{T}}\boldsymbol{\beta}_0 + \eta_{\boldsymbol{\beta}_0}(t_i))\bigg[\mathbf{x}_i + \frac{\partial}{\partial\boldsymbol{\beta}}\eta_{\boldsymbol{\beta}}(t_i)\bigg|_{\boldsymbol{\beta}=\boldsymbol{\beta}_0}\bigg] w_2(\mathbf{x}_i).$$

Using the fact that $\eta_{\boldsymbol{\beta}_0} = \eta_0$ and noting that N5 implies $E[\Psi(y_i, \mathbf{x}_i^{\mathrm{T}}\boldsymbol{\beta}_0 + \eta_{\boldsymbol{\beta}_0}(t_i))|(\mathbf{x}_i, t_i)] = 0$, it follows that $L_n$ is asymptotically normally distributed with covariance matrix $\boldsymbol{\Sigma}$. Therefore, it remains to show that $L_n - \hat{L}_n \xrightarrow{p} 0$.

We have the expansion $\hat{L}_n - L_n = L_n^1 + L_n^2 + L_n^3 + L_n^4$, where

$$L_n^1 = n^{-1/2}\sum_{i=1}^{n} \chi(y_i, \mathbf{x}_i^{\mathrm{T}}\boldsymbol{\beta}_0 + \eta_0(t_i))\bigg[\mathbf{x}_i + \frac{\partial}{\partial\boldsymbol{\beta}}\eta_{\boldsymbol{\beta}}(t_i)\bigg|_{\boldsymbol{\beta}=\boldsymbol{\beta}_0}\bigg] w_2(\mathbf{x}_i)\hat{v}_0(t_i),$$

$$L_n^2 = n^{-1/2} \sum_{i=1}^n \Psi(y_i, \mathbf{x}_i^{\mathrm{T}} \boldsymbol{\beta}_0 + \eta_{\boldsymbol{\beta}_0}(t_i)) w_2(\mathbf{x}_i) \hat{\mathbf{v}}_0(t_i),$$

$$L_n^3 = n^{-1} \sum_{i=1}^n \chi(y_i, \mathbf{x}_i^{\mathrm{T}} \boldsymbol{\beta}_0 + \eta_0(t_i)) w_2(\mathbf{x}_i)(n^{1/4} \hat{\mathbf{v}}_0(t_i))(n^{1/4} \hat{v}_0(t_i)),$$

$$L_n^4 = (2n)^{-1} \sum_{i=1}^n \chi_1(y_i, \mathbf{x}_i^{\mathrm{T}} \boldsymbol{\beta}_0 + \xi(t_i)) \left[\mathbf{x}_i + \frac{\partial}{\partial \boldsymbol{\beta}} \eta_{\boldsymbol{\beta}}(t_i)\bigg|_{\boldsymbol{\beta}=\boldsymbol{\beta}_0}\right] w_2(\mathbf{x}_i)(n^{1/4} \hat{v}_0(t_i))^2,$$

with $\hat{v}_0(t) = \hat{\eta}_{\boldsymbol{\beta}_0}(t) - \eta_0(t)$, $\hat{\mathbf{v}}_0(t) = (\hat{v}_{1,0}(t), \ldots, \hat{v}_{p,0}(t))^{\mathrm{T}} = \partial \hat{v}(\boldsymbol{\beta}, t)/\partial \boldsymbol{\beta}|_{\boldsymbol{\beta}=\boldsymbol{\beta}_0}$ defined as in (A.6), $\hat{v}$ defined as in (A.5) and $\xi(t_i)$ an intermediate point between $\hat{\eta}_{\boldsymbol{\beta}_0}(t_i)$ and $\eta_0(t_i)$. It is easy to see that $L_n^3 \xrightarrow{p} 0$ and $L_n^4 \xrightarrow{p} 0$ follow from N1(c) and N2. To complete the proof, it remains to show that $L_n^j \xrightarrow{p} 0$ for $j = 1, 2$, which will follow from N1(c)–(e) and N5 using entropy arguments similar to those considered in [3]. Details can be found in [8]. $\square$

**Acknowledgments.** We wish to thank the Associate Editor and two anonymous referees for valuable comments which led to an improved version of the original paper.

## REFERENCES

[1] BIANCO, A. and BOENTE, G. (1996). Robust nonparametric generalized regression estimation. Impresiones Previas del Departamento de Matemática, FCEN, September 1996.

[2] BIANCO, A. and BOENTE, G. (2002). On the asymptotic behavior of one-step estimates in heteroscedastic regression models. *Statist. Probab. Lett.* **60** 33–47. MR1945676

[3] BIANCO, A. and BOENTE, G. (2004). Robust estimators in semiparametric partly linear regression models. *J. Statist. Plann. Inference* **122** 229–252. MR2057924

[4] BIANCO, A., GARCÍA BEN, M. and YOHAI, V. (2005). Robust estimation for linear regression with asymmetric errors. *Canad. J. Statist.* **33** 511–528. MR2232377

[5] BIANCO, A. and YOHAI, V. (1995). Robust estimation in the logistic regression model. *Lecture Notes in Statist.* **109** 17–34. Springer, New York. MR1491394

[6] BOENTE, G. and FRAIMAN, R. (1991). Strong order of convergence and asymptotic distribution of nearest neighbor density estimates from dependent observations. *Sankhyā Ser. A* **53** 194–205. MR1180044

[7] BOENTE, G., FRAIMAN, R. and MELOCHE, J. (1997). Robust plug-in bandwidth estimators in nonparametric regression. *J. Statist. Plann. Inference* **57** 109–142. MR1440232

[8] BOENTE, G., HE, X. and ZHOU, J. (2005). Robust estimates in generalized partially linear models (with full appendix). Technical report, available at http://www.ic.fcen.uba.ar/preprints/boentehezhou.pdf.

[9] CANTONI, E. and RONCHETTI, E. (2001). Robust inference for generalized linear models. *J. Amer. Statist. Assoc.* **96** 1022–1030. MR1947250

[10] CANTONI, E. and RONCHETTI, E. (2001). Resistant selection of the smoothing parameter for smoothing splines. *Statist. Comput.* **11** 141–146. MR1837133

ROBUST SEMIPARAMETRIC REGRESSION 23

[11] Carroll, R., Fan, J., Gijbels, I. and Wand, M. (1997). Generalized partially linear single-index models. *J. Amer. Statist. Assoc.* **92** 477–489. MR1467842
[12] Croux, C. and Haesbroeck, G. (2002). Implementing the Bianco and Yohai estimator for logistic regression. *Comput. Statist. Data Anal.* **44** 273–295. MR2020151
[13] Gao, J. and Shi, P. (1997). M-type smoothing splines in nonparametric and semiparametric regression models. *Statist. Sinica* **7** 1155–1169. MR1488663
[14] He, X., Fung, W.K. and Zhu, Z.Y. (2005). Robust estimation in generalized partial linear models for Clustered Data. *J. Amer. Statist. Assoc.* **100** 1176–1184. MR2236433
[15] He, X., Zhu, Z. and Fung, W. (2002). Estimation in a semiparametric model for longitudinal data with unspecified dependence structure. *Biometrika* **89** 579–590. MR1929164
[16] Huber, P. (1981). *Robust Statistics*. Wiley, New York. MR0606374
[17] Künsch, H., Stefanski, L. and Carroll, R. (1989). Conditionally unbiased bounded-influence estimation in general regression models with applications to generalized linear models. *J. Amer. Statist. Assoc.* **84** 460–466. MR1010334
[18] Leung, D., Marrot, F. and Wu, E. (1993). Bandwidth selection in robust smoothing. *J. Nonparametr. Statist.* **4** 333–339. MR1256384
[19] McCullagh, P. and Nelder, J.A. (1989). *Generalized Linear Models*. Chapman and Hall, London. MR0727836
[20] Pollard, D. (1984). *Convergence of Stochastic Processes*. Springer, New York. MR0762984
[21] Severini, T. and Staniswalis, J. (1994). Quasi-likelihood estimation in semiparametric models. *J. Amer. Statist. Assoc.* **89** 501–511. MR1294076
[22] Severini, T. and Wong, W. (1992). Generalized profile likelihood and conditionally parametric models. *Ann. Statist.* **20** 1768–1802. MR1193312
[23] Stefanski, L., Carroll, R. and Ruppert, D. (1986). Bounded score functions for generalized linear models. *Biometrika* **73** 413–424. MR0855901
[24] Wang, F. and Scott, D. (1994). The $L_1$ method for robust nonparametric regression. *J. Amer. Statist. Assoc.* **89** 65–76. MR1266287

Graciela Boente
Departamento de Matemática
e Instituto de Cálculo
Facultad de Ciencias Exactas y Naturales
Ciudad Universitaria, Pabellón 2
Buenos Aires, C1428EHA
Argentina
E-mail: gboente@dm.uba.ar

Xuming He
Department of Statistics
University of Illinois at Urbana-Champaign
725 South Wright Street
Champaign, Illinois 61820
USA
E-mail: x-he@uiuc.edu

Jianhui Zhou
Department of Statistics
University of Virginia
Charlottesville, Virginia 22904
USA
E-mail: jz9p@virginia.edu